%% file: main.tex
\documentclass[journal]{IEEEtran}

\usepackage{amsmath}
\usepackage{graphicx}
\usepackage{float} 
\usepackage{amssymb}
\usepackage{subfig}
\usepackage{array}
\usepackage{tikz}
\usepackage{xcolor}
\usepackage{pgfplots}
\usepackage{booktabs}
\usepackage[acronym,shortcuts]{glossaries}
\usepackage[normalem]{ulem}
\usepackage{orcidlink}
\usepackage{hyperref}
\usepackage{balance}

\hypersetup{hidelinks}
\usepgfplotslibrary{groupplots} 
\pgfplotsset{compat=1.18} 

\input{acro_list} 

\usetikzlibrary{positioning, fit, backgrounds}
\tikzset{
    block/.style = {draw, rectangle, minimum height=0.9cm, minimum width=1.1cm, align=center, fill=white, font=\scriptsize, inner sep=2pt},
    tallblock/.style = {draw, rectangle, minimum height=2.4cm, minimum width=0.8cm, align=center, fill=white, font=\scriptsize, inner sep=2pt},
    arrow/.style = {-{Latex[length=1.5mm, width=1.2mm]}},
    stagebox/.style = {fill=gray!15, inner sep=3pt, rounded corners=2pt}
}

\begin{document}

\title{Comparative Analysis of Filter Bank Waveforms for ISAC Systems: 2D-FFT FB and AFBM}

\author{
\IEEEauthorblockN{Polianna~Basso\IEEEauthorrefmark{1}\textsuperscript{\orcidlink{0009-0000-2202-7464}},~\IEEEmembership{Student Member,~IEEE}, Victor~Augusto~Del~Monego\IEEEauthorrefmark{1}\textsuperscript{\orcidlink{0009-0001-6817-3093}},
Hyeon~Seok~Rou\IEEEauthorrefmark{2}\textsuperscript{\orcidlink{0000-0003-3483-7629}},~\IEEEmembership{Member,~IEEE},
Bruno~S.~Chang\IEEEauthorrefmark{1}\textsuperscript{\orcidlink{0000-0003-0232-7640}},~\IEEEmembership{Member,~IEEE},
Giuseppe~Thadeu~Freitas~de~Abreu\IEEEauthorrefmark{2}\textsuperscript{\orcidlink{0000-0002-5018-8174}},~\IEEEmembership{Senior Member,~IEEE}
and Didier~Le~Ruyet\IEEEauthorrefmark{3}\textsuperscript{\orcidlink{0000-0002-9673-2075}},~\IEEEmembership{Senior Member,~IEEE}}

\vspace{0.7em}

\IEEEauthorblockA{\IEEEauthorrefmark{1}CPGEI/Electronics Department, Federal University of Technology - Paran\'a, Curitiba, Brazil\\
\IEEEauthorrefmark{2}School of Computer Science and Engineering, Constructor University, Bremen, Germany\\
\IEEEauthorrefmark{3}CEDRIC/Conservatoire National des Arts et M\'etiers - Paris, France}\\[0.15ex]
Emails: \{poliannabasso, victormonego\}@alunos.utfpr.edu.br,  bschang@utfpr.edu.br, \{hrou, gabreu\}@constructor.university, didier.le\_ruyet@cnam.fr%
}

\IEEEaftertitletext{\vspace{-1\baselineskip}}

\maketitle

%
%

\begin{abstract}
We present a comparative analysis of two promising filter bank multicarrier waveforms for \ac{ISAC} systems operating in high-mobility environments. In particular, we consider \ac{2DFFTFB} modulation and the recently-proposed \ac{AFBM} waveforms, which exploit different signal representations across the time-frequency, delay-Doppler and affine domains to improve robustness against Doppler effects, each resulting in distinct features.
On the one hand, \ac{2DFFTFB} improves spectral localization under high Doppler conditions, while \ac{AFBM} combines chirp-based modulation with filter bank processing to achieve low \ac{PAPR} and reduced \ac{OOBE}.
These schemes are compared in terms of \ac{OOBE}, \ac{PAPR} and \ac{BER}, as well as the delay and Doppler cuts of the \ac{AF} under deterministic and random sensing symbols. Results for \ac{OTFS} and \ac{AFDM} are also included to provide additional basis of comparison.
The results show the performance trade-offs between the two \ac{FB}-based schemes, providing insights into waveform candidates for high-mobility \ac{ISAC} scenarios such as vehicular communications and future \ac{6G} networks.

\end{abstract}


\IEEEpeerreviewmaketitle

\vspace{-2ex}
\section{Introduction}

In an increasingly connected world, vehicular networks have become key to safer and more reliable mobility, with \ac{V2X} communication supporting both communication and sensing functionalities. Traditionally, these functions are implemented separately, leading to inefficient use of hardware and spectrum resources \cite{Cong2023}. Integrated sensing and communications (ISAC) has emerged as a promising paradigm that integrates both over shared resources, particularly in high-dynamic vehicular scenarios, combining radar sensing and wireless communications into a unified framework that improves spectral efficiency and reduces hardware redundancy \cite{Du2023}.
One important application is predictive safety, where vehicles combine onboard and external information exchanged with nearby vehicles and \acp{RSU} to anticipate hazards, extending their perception of the environment and improving traffic safety \cite{Cheng2022}.
The deployment of such technology remains challenging, however, due to the severe delay and Doppler spreads of doubly-dispersive channels, which stem from the relative motion between transmitters, receivers and surrounding objects and result in channel variations and performance degradation \cite{Rou2024}.

Waveform design is thus central to enabling reliable sensing and communication in high-mobility environments \cite{Rou2024}. Conventional orthogonal frequency division multiplexing (OFDM) offers high spectral efficiency and simple equalization, but suffers from inter-carrier interference under strong Doppler effects \cite{OFDM}. Filter bank multi-carrier (FBMC) improves spectral localization and reduces out-of-band emissions (OOBE) through subcarrier filtering, at the cost of increased transceiver complexity \cite{FBMC}. Orthogonal time-frequency space (OTFS) operates in the delay-Doppler domain and is highly robust to doubly dispersive channels, but its OFDM-based structure still yields considerable OOBE \cite{OTFS}.

More recently, chirp-domain waveforms such as affine frequency division multiplexing (AFDM) have attracted attention for their robustness to Doppler dispersion in high-mobility scenarios \cite{AFDM}, while inheriting the high peak-to-average power ratio (PAPR) and limited spectral containment of OFDM-based structures. While \ac{OTFS} relies on embedded pilots with sizeable guard regions and incurs pilot overhead, \ac{AFDM}'s tunable chirp parameters enable single-pilot, lower-complexity channel estimation \cite{AFDM}. 
Motivated by these challenges, this paper compares two waveform structures for high-mobility ISAC: Two-Dimensional FFT Precoded Filter Bank (2D-FFT FB) \cite{Junior2024} and Affine Filter Bank Modulation (AFBM) \cite{AFBM}. Both share a filter bank core and differ only in the precoding stage, so comparing them, which has not yet been done under a common framework, isolates the effect of the precoder itself. In turn, \ac{OTFS} and \ac{AFDM} are added as \ac{OFDM}-based baselines, allowing the filter bank core and the precoding stage to be assessed separately. The waveforms are evaluated under doubly-dispersive channel conditions in terms of \ac{BER}, \ac{PAPR}, \ac{OOBE}, the delay and Doppler cuts of the \ac{AF} \cite{RouAF2026}, and the computational complexity, to assess their trade-offs for high-mobility ISAC applications.

%

\textit{Notation:} Bold lowercase and uppercase letters denote vectors and matrices. The operators $(\cdot)^T$ and $(\cdot)^H$ denote transpose and Hermitian transpose, while $\operatorname{diag}(\cdot)$, $\operatorname{vec}(\cdot)$ and $\otimes$ denote diagonalization, vectorization and the Kronecker product. $\mathbf{I}_Q$ denotes the $Q\times Q$ identity matrix, and $j=\sqrt{-1}$. The unitary $Q$-point discrete Fourier transform (DFT) matrix is denoted by $\mathbf{F}_Q$, with $\mathbf{F}_Q^H$ representing the  inverse DFT. 

%
%
\begin{figure*}[!t]
    \centering
    \includegraphics[width=\textwidth]{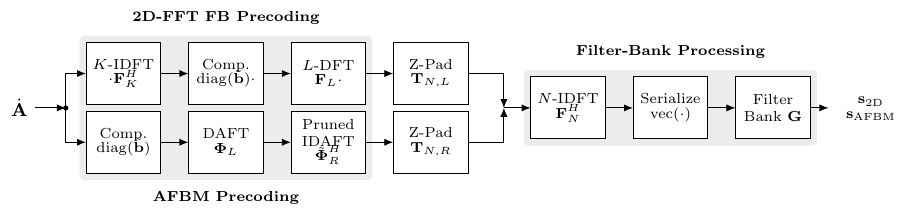}
    \caption{Block diagram of the transmission chain of the considered waveforms. The upper branch shows the 2D-FFT filter-bank transmitter, while the lower branch shows the AFBM transmitter, including subcarrier allocation, compensation, affine-domain spreading, frequency-domain zero padding, inverse transforms, and prototype filtering.}
    \label{fig:transmission_chains}
\end{figure*}

\vspace{-2ex}
\section{Waveform Transmission}
\label{sec:transmission}

This section aims to describe the compared waveforms' modulation and transmission schemes. To ensure a consistent comparison paradigm, all considered waveform transmission schemes, shown in Fig. \ref{fig:transmission_chains}, originate from the exact same data payload.

Let $\mathbf{x} \in \mathcal{D}^{N_d \times 1}$ denote the common vector of $N_d$ transmitted complex symbols, where $\mathcal{D}$ represents the Quadrature Amplitude Modulation (QAM) alphabet.
Both schemes are configured with $L$ signal dimensions per frame. In the following formulations, we denote the number of time symbols for filter both schemes as {$K = 2N_d/L$}.

\vspace{-1ex}
\subsection{The 2D-FFT FB Transmitter in Brief}\label{2dfft}

The \ac{2DFFTFB} scheme, proposed in~\cite{2DFFT_FB}, can be interpreted as a variation of OTFS~\cite{OTFS} in which the OFDM kernel is replaced by a filter bank multicarrier structure. While OFDM ensures complex orthogonality with a time-frequency lattice density $1/(TF) \leq 1$~\cite{lattice}, FBMC/OQAM operates at $TF = 0.5$ at the cost of intrinsic interference. To restore complex orthogonality, 2D-FFT FB spreads $L/2$ complex symbols across the $L$ subcarriers of each symbol interval through DFT-based precoding and filter compensation, so that the equivalent spacing returns to $TF = 1$ without loss of information rate.
Based on this, the symbol matrix $\mathbf{A} \in \mathcal{D}^{(L/2) \times K}$ such that $\mathbf{x} = \mathrm{vec}(\mathbf{A})$ is defined. This symbol matrix is mapped into the data matrix $\mathbf{\dot{A}} \in \mathbb{C}^{L \times K}$ as follows:
%
%
\vspace{-1ex}
\begin{equation}
\mathbf{\dot{A}}= \bar{\mathbf{\Xi}} \mathbf{A}
\vspace{-1ex}
\end{equation}
where
\begin{equation}
\bar{\mathbf{\Xi}} =
\begin{bmatrix}
\mathbf{I}_{L/4} & \mathbf{0}_{L/4} \\
\mathbf{0}_{L/2} & \mathbf{0}_{L/2} \\
\mathbf{0}_{L/4} & \mathbf{I}_{L/4}
\end{bmatrix}.
\end{equation}

The resulting matrix contains $L/2$ active delay domain bins and $L/2$ zero-valued guard bins along its first axis. Its second axis contains $K$  bins. The guard bins mitigate interference introduced by the filter bank. The grid is subsequently mapped to the time-frequency domain signal at the input of the filter bank $\mathbf{X} \in \mathbb{C}^{L \times K}$, expressed as
\vspace{-0.5ex}
\begin{equation}
\mathbf{X} = \mathbf{F}_{L} \textrm{diag}\{\tilde{\mathbf{b}}\} \mathbf{\dot{A}}\mathbf{F}^H_{K},
\label{OTFSFB}
\end{equation}
where  $\tilde{\mathbf{b}}$ is the vector representing the filter compensation coefficients.

In the filter bank implementation, the time-domain signal is obtained by applying 
an extended $N$-point IDFT to the time-frequency matrix $\mathbf{X}$. This operation 
results in the matrix $\mathbf{D} \in \mathbb{C}^{N \times K}$, defined as
\vspace{-0.5ex}
\begin{equation}
    \mathbf{D} = \mathbf{{F}}^H_{N} \mathbf{T}_{N,L}\mathbf{X},
\end{equation}
where $\mathbf{T}_{NL} \in \mathbb{R}^{N \times L}$ represents the frequency-domain zero-padding matrix, with $N > L$. Consequently, the signal undergoes a temporal expansion, where each column of $\mathbf{D}$ corresponds to a time-domain sequence associated with a given symbol interval.

At this stage, the system operates with dimension $N$ instead of $L$, which implies that the prototype filter is defined with length $ON$, where $O$ denotes the overlap factor of the prototype filter. After conversion to the time domain, each vector ${d}_{k}$ is filtered by the prototype filter.  To model the filtering operation a set of diagonal matrices $\mathbf{G}_\rho$ associated with the prototype filter coefficients is defined and arranged diagonally along a block Toeplitz matrix $\mathbf{G} \in \mathbb{R}^{M \times NK}$, where $M = ON + (K - 1)\frac{N}{2}$\cite{2DFFT_FB}. 

In this manner, each block corresponds to a time-shifted version of the prototype filter. 
Thus, the filter bank output can be represented by $\mathbf{s_{2D}} \in \mathbb{C}^{M \times 1}$, given by
\begin{equation}
\mathbf{s_{2D}} = \mathbf{G}\mathbf{d},
\end{equation}
where $\mathbf{d} = \mathrm{vec}(\mathbf{D}) \in \mathbb{C}^{NK \times 1}$.

\vspace{-2ex}
\subsection{The AFBM Transmitter in Brief}

AFBM combines affine-domain spreading with filter bank multicarrier processing in order to provide robustness against doubly dispersive channels while simultaneously reducing OOBE and PAPR \cite{AFBM}. The scheme can be interpreted as an extension of AFDM in which the conventional chirp-based subcarriers are replaced by filtered chirp subcarriers generated through a filter bank structure.

Consider a system employing $L$ subcarriers and $K$ time symbols. In order to preserve near complex orthogonality, only $L/2$ subcarriers are effectively used for data transmission, similarly to 2D-FFT FB. The  QAM symbol vector is first mapped into the time-frequency grid $\mathbf{a} \in \mathbb{C}^{LK\times1}$ through
\vspace{-0.5ex}
\begin{equation}
\mathbf{a\triangleq vec(\dot{A})} = (\mathbf{I}_{K} \otimes \bar{\mathbf{\Xi}})\mathbf{x},
\end{equation}
where $\mathbf{\dot{A}}$ and $\mathbf{x}$ are as defined in Section~\ref{2dfft}, and $\bar{\mathbf{\Xi}}\in \mathbb{C}^{L\times \frac{L}{2}}$ denotes the allocation matrix presented in (2). 

After symbol allocation, a compensation stage combined with affine-domain spreading is applied in order to restore near complex orthogonality. 
The vectorized form of the DAFT-spread transmit signal $\mathbf{g} \in \mathbb{C}^{LK\times 1}$ is expressed as 
\vspace{-1ex}
\begin{equation}
\mathbf{g} = \mathbf{C}  \mathbf{a}.
\vspace{-0.5ex}
\end{equation}

The spreading matrix $\mathbf{C} \in \mathbb{C}^{LK\times LK}$ is defined as
\vspace{-1ex}
\begin{equation}
\mathbf{C}
=
\mathbf{I}_{K} \otimes
\left(
\boldsymbol{\Phi}_L
\mathrm{diag}(\tilde{\mathbf{b}})
\right),
\vspace{-0.5ex}
\end{equation}
where $\mathrm{diag}(\tilde{\mathbf{b}})$ is as defined in Section~\ref{2dfft}, and $\boldsymbol{\Phi}_L \in \mathbb{C}^{L\times L}$ represents the $L$-point \ac{DAFT} matrix, defined as
\vspace{-1ex}
\begin{equation}
\mathbf {\Phi}_L
\triangleq
\mathbf{\Lambda}_{c_1,L}
\mathbf{F}_L
\mathbf{\Lambda}_{c_2,L},
\vspace{-0.5ex}
\end{equation}
where $\mathbf{\Lambda}_{c_1,L} \in \mathbb{C}^{L\times L}$ and  $\mathbf{\Lambda}_{c_2,L} \in \mathbb{C}^{L\times L}$ are diagonal chirp modulation matrices associated with the affine parameters $c_1$ and $c_2$ given by
\vspace{-1ex}
\begin{equation}
\mathbf{\Lambda}_{c_i,L} = \mathrm{diag}\big[\, e^{-j2\pi c_i n^{2}} \,\big]_{n=0}^{L-1} \in \mathbb{C}^{L\times L}.
\vspace{-0.5ex}
\end{equation}

These chirp parameters are derived from the channel according to the orthogonality condition below
\begin{equation}
2(\mathbf{\nu^{\max}}+\mathbf{\xi})(\mathbf{\tau^{\max}}+1)+\mathbf{\tau^{\max}}\leq \mathbf{N},
\vspace{-0.5ex}
\end{equation}
where $\nu^{\max}$ and $\tau^{\max}$ represent the channel's maximum normalized Doppler shift and delay, respectively, and $\xi$ is a guard width, used to indicate the number of reserved additional  elements to account for Doppler interference leakage. 

Next, the affine-precoded symbols are converted to the time domain via an extended inverse \ac{DAFT} operation combined with frequency-domain zero padding.
The $R$-point transform is larger than the $L$-dimensional affine-spread block. Therefore, with $L < R < N$, its pruned version can be defined as
\vspace{-0.5ex}
\begin{equation}
\tilde{\boldsymbol{\Phi}}_{R}
\triangleq
\begin{bmatrix}
\mathbf{I}_{L} & \mathbf{0}_{L\times(R-L)}
\end{bmatrix}
\boldsymbol{\Phi}_{R}
\in\mathbb{C}^{L\times R}.
\vspace{-0.5ex}
\end{equation}

The inverse DAFT processing stage is represented by $\mathbf{Q}\in\mathbb{C}^{NK\times LK}$, defined as
\vspace{-1ex}
\begin{equation}
\mathbf{Q}
\triangleq \mathbf{I}_{K} \otimes \mathbf{Q}_R =
\mathbf{I}_{K} \otimes (\mathbf{F}_N^H
\mathbf{T_{N,R}}
\tilde{\boldsymbol{\Phi}}_R^H),
\vspace{-0.5ex}
\end{equation}
where $\tilde{\boldsymbol{\Phi}}_R \in \mathbb{C}^{L\times R}$ denotes the pruned \ac{DAFT} matrix. Therefore, the time-domain signal before filtering is obtained as
\vspace{-1ex}
\begin{equation}
\tilde{\mathbf{f}}
= \mathbf{Q}\mathbf{g}
= \mathbf{QC}\mathbf{a}.
\vspace{-1ex}
\end{equation}

The filtering operation can be represented by a block Toeplitz matrix $\mathbf{G}$, the same way as seen in Section~\ref{2dfft}. Finally, the transmitted \ac{AFBM} signal $\mathbf{s_{AFBM}} \in \mathbb{C}^{M\times1}$ can be written as
\vspace{-1ex}
\begin{equation}
\mathbf{s_{AFBM}} = \mathbf{G}\tilde{\mathbf{f}} = \mathbf{G}\mathbf{Q}\mathbf{C}\mathbf{a}.
\vspace{-1ex}
\end{equation}

Since both FB-based schemes rely on a well-localized prototype filter, the channel-induced interference is confined by the filter itself, so neither requires a cyclic prefix (CP), unlike OTFS and AFDM.
Accordingly, the channel in Section III is modeled as a linear convolution through $\boldsymbol{\Pi}$.

\section{Channel Model}
\label{sec:channel}

In multipath channels, the received signal is formed by the superposition of multiple copies of the transmitted signal, each affected by different delays and Doppler shifts.
For this paper, a doubly-dispersive channel is considered, arising 
from multipath propagation and relative motion between the transmitter and receiver \cite{Rou2024}.

The channel is modeled as a superposition of $P$ propagation paths, where each path 
is defined by a complex gain $h_p \in \mathbb{C}$, a delay $\tau_p$, and a Doppler 
shift $\nu_p$.
The corresponding time-varying impulse response \cite{Rou2024} can be 
expressed as
\begin{equation}
h(t,\tau) = \sum_{p=1}^{P} h_p \, e^{j 2\pi \nu_p t} \delta(\tau - \tau_p),
\end{equation}
where $t$ denotes time and $\tau$ represents the delay variable.

This model captures the time-frequency selectivity of high-mobility channels. By discretizing the continuous-time model and assuming a finite observation window,  the channel can be represented in matrix form. Defining the normalized integer delay 
$\ell_p \triangleq \lfloor \tau_p / T_s \rceil$ and the normalized digital Doppler 
shift $f_p \triangleq M \nu_p / f_s$, with $T_s$ denoting the sampling period and 
$f_s \triangleq 1/T_s$ the sampling frequency, the equivalent discrete-time channel 
matrix $\mathbf{H} \in \mathbb{C}^{M \times M}$ can be expressed as \cite{Rou2024}
\begin{equation}
\mathbf{H} = \sum_{p=1}^{P} h_p \, \mathbf{Z}^{f_p} \mathbf{\Pi}^{\ell_p},
\end{equation}
where $\mathbf{\Pi} \in \mathbb{R}^{M \times M}$ is the linear shift matrix, 
so that $\mathbf{\Pi}^{\ell_p}$ applies a delay of $\ell_p$ samples, and 
$\mathbf{Z} \triangleq \mathrm{diag}[ e^{-j2\pi \frac{0}{M}}, \dots, 
e^{-j2\pi \frac{M-1}{M}} ] \in \mathbb{C}^{M \times M}$ introduces the 
Doppler-induced phase shifts.
The same channel model is applied for all evaluated waveforms.

\begin{figure*}[!t]
    \centering
    \includegraphics[width=\textwidth]{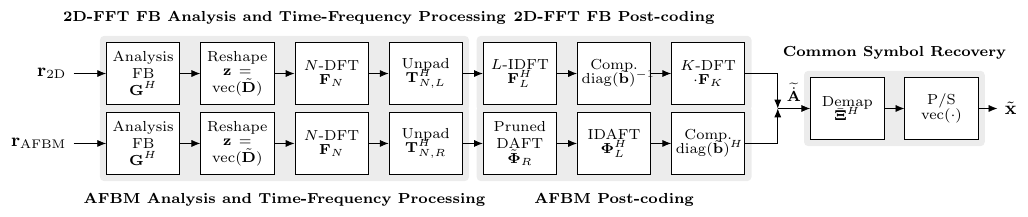}
    \caption{Block diagram of the reception chains of the considered waveforms. The upper branchshows the 2D-FFT filter-bank receiver, and the lower branch shows the AFBM receiver, including analysis filtering, affine-domain processing, compensation, and recovery of the transmitted QAM symbols.}
    \label{fig:reception_chains}
\end{figure*}

\section{Waveform Reception}
\label{sec:reception}

The received signal vector after propagation through the channel can be expressed as
\begin{equation}
\mathbf{r} = \mathbf{H}\mathbf{s}+\mathbf{n},
\end{equation}
where $\mathbf{n} \sim \mathcal{CN}(\mathbf{0}, \sigma_n^2 \mathbf{I}_M)$ is a complex \ac{AWGN} vector, whose dimension corresponds to the length of the transmitted signal.

\subsection{The 2D-FFT FB Receiver in Brief}

The receiver performs the inverse operations of the transmitter, as
shown in Fig.~\ref{fig:reception_chains}. First, the received signal $\mathbf{z}\in\mathbb{C}^{NK\times1}$ is demodulated by the analysis filter bank, resulting in
\begin{equation}
    \mathbf{z}
    =
    \mathbf{G}^{H}\mathbf{r}_{\mathrm{2D}}.
\end{equation}

By splitting the vector $\mathbf{z}$ into $K$ columns, we obtain the received matrix $\tilde{\mathbf{D}}\in\mathbb{C}^{N\times K}$.
After DFT processing and removal of the frequency-domain zero padding, the intermediate time-frequency matrix $\tilde{\mathbf{X}}\in\mathbb{C}^{L\times K}$ is recovered as
\begin{equation}
    \tilde{\mathbf{X}}
    =
   \mathbf{T}_{N,L}^{H}\mathbf{F}_N{\tilde{\mathbf{D}}}.
\end{equation}

Recalling the time-frequency mapping in~\eqref{OTFSFB}, the transmitter generates $\mathbf{X}$ from $\dot{\mathbf{A}}$ through three successive operations: filter compensation by $\operatorname{diag}\{\tilde{\mathbf{b}}\}$, frequency spreading by the $L$-point DFT along the first axis, and the $K$-point IDFT along the second axis.
The receiver therefore reverses these operations in the opposite order. Since $\mathbf{F}_L$ and $\mathbf{F}_K$ are unitary, their inverses are given by their Hermitian transposes, so left-multiplying $\tilde{\mathbf{X}}$ by $\mathbf{F}_L^H$ despreads the symbols across the subcarriers, while right-multiplying it by $\mathbf{F}_K$ undoes the transform along the time axis.

Finally, since the compensation matrix is diagonal with nonzero entries, its effect is removed by its inverse, yielding the estimate of the data matrix
\begin{equation}
    \widetilde{\dot{\mathbf{A}}}
    =
    \operatorname{diag}(\tilde{\mathbf{b}})^{-1}
    \mathbf{F}_{L}^{H}
    \tilde{\mathbf{X}}
    \mathbf{F}_{K}.
\end{equation}

\vspace{-2ex}
\subsection{The AFBM Receiver in Brief}

At the receiver, the inverse operations of the AFBM transmitter are
performed, as shown in Fig.~\ref{fig:reception_chains}.
First, the
received signal is processed by the analysis filter bank and
affine-domain demodulator:
\vspace{-0.5ex}
\begin{equation}
    \mathbf{z}
    =
    (\mathbf{G}\mathbf{Q})^{H}
    \mathbf{r}_{\mathrm{AFBM}},
    \vspace{-0.5ex}
\end{equation}
where $\mathbf{z}\in\mathbb{C}^{LK\times1}$ denotes the received vector
before affine-domain despreading and compensation.
The estimated allocated-grid vector is then obtained as
\vspace{-0.5ex}
\begin{equation}
    \tilde{\mathbf{a}}
    =
    \mathbf{C}^{H}\mathbf{z}.
\vspace{-0.5ex}
\end{equation}
where $\mathbf{\tilde{a} = vec(\tilde{\dot{A}})}$. 

For both waveforms, the estimated data-symbol matrix is then recovered by reversing the subcarrier allocation operation. After that, parallel-to-serial conversion yields the estimated transmitted
symbol vector by 
\vspace{-0.5ex}
\begin{equation}
    \tilde{\mathbf{x}} = \operatorname{vec}\left(\tilde{\mathbf{A}}\right) = \operatorname{vec}\left({\bar{\mathbf{\Xi}}^{H}\widetilde{\dot{\mathbf{A}}}}\right).
\vspace{-0.5ex}
\end{equation}

\vspace{-1ex}
\section{Ambiguity Function}

The sensing capability of each waveform is assessed through its discrete \ac{AF} ~\cite{Rou2024}, which for a transmit signal $\mathbf{s} \in \mathbb{C}^{M\times 1}$ is
\vspace{-0.5ex}
\begin{equation}
A[\ell, f] = \sum_{n=0}^{M-1} s[n]\, s^{*}[n-\ell]\, e^{-j2\pi \frac{f n}{M}},
\vspace{-0.5ex}
\end{equation}
where $\ell$ and $f$ are the delay and Doppler indices, and $s[n-\ell] = 0$ outside $[0, M-1]$, since no CP is used.
To obtain a system-agnostic representation, normalized coordinates $\tau_{\mathrm{norm}} = \ell/M \in [-1, 1]$ and $\nu_{\mathrm{norm}} = f/M \in [-0.5, 0.5]$ are adopted. The delay and Doppler resolutions are then assessed through the zero-Doppler cut $A_{\tau}[\ell] = A[\ell, 0]$ and the zero-delay cut $A_{\nu}[f] = A[0, f]$, both normalized to unit peak.

\vspace{-2ex}
\section{Results and discussion}\label{sec:results}


This section presents a comparative performance evaluation of the \ac{2DFFTFB} and \ac{AFBM} schemes, against \ac{OTFS} and \ac{AFDM} as baselines, under high-mobility vehicular channel conditions. To ensure a fair comparison, in each transmission interval for all waveforms $N_d = 1024$ complex symbols ($L \times K = 256 \times 8$ for filter-bank based systems, with $N = 512$, $R=272$ for AFBM and $32\times32$ for OTFS) were transmitted at a carrier frequency of 4 GHz, with 20 MHz of bandwidth.
The considered velocity of 500~km/h follows the IMT-2020 mobility requirement~\cite{itur_m2410_2017} and represents a worst-case Doppler condition. The analysis is based on Monte Carlo simulations where all waveforms are generated under identical system configurations to ensure a fair analysis. 
The channel follows a doubly dispersive Vehicular A \cite{itur_m1225_1997}  model with multipath and Doppler effects, as described in Section \ref{sec:channel}.  Filter bank based waveforms use the Hermite and PHYDYAS prototype filter models \cite{Haas1997, phydias}.
To ensure a fair comparison among the considered waveforms, a MMSE equalizer was used for all waveforms. In particular, AFBM employs a hybrid time-domain/filter-bank equalization approach \cite{AFBMSIR}. For BER and random QAM symbol AF simulations, 1000 frames were compiled. OOBE and PAPR simulations were conducted using 10000 frames. The unitary input AF is computed by a single frame simulation. In terms of chirp matrices, AFBM chirps were defined as $c_1 = \frac{2(\nu^{max}+\xi)+1}{2*L}$, $\lambda_1 = \frac{2(\nu^{max}+\xi)+1}{2*L}$ and $c_2 = \lambda_2 = \pi$. AFDM chirps were defined as $c_1 = \frac{2(\nu^{max}+\xi)+1}{2*L} $ and $c_2 = \frac{1}{L^2\pi}$. 

%
%

\begin{figure}[tb]
    \centering
    \includegraphics[width=\columnwidth]{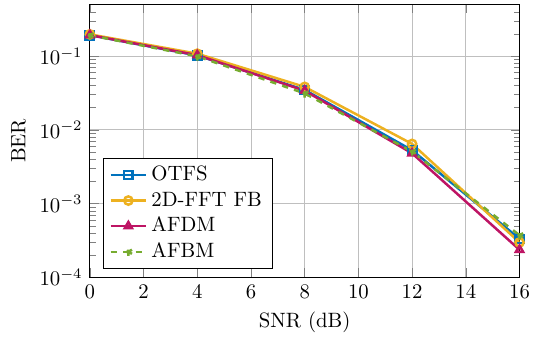}
    \vspace{-1.5em}
    \caption{\ac{BER} performance versus \ac{SNR} for \ac{AFBM}, \ac{AFDM}, 2D-FFT FB, and \ac{OTFS}, on a Vehicular A channel with a maximum user velocity of 500 km/h.}
    \label{fig:BER}
\end{figure}

Starting with \ac{BER} performance, the results in Fig.~\ref{fig:BER} show that all four waveforms are robust in doubly-selective channel conditions and achieve similar performance, with differences of approximately below 1 dB across the entire simulated \ac{SNR} range. 

\begin{figure}[htbp]
    \centering
    \includegraphics[width=\columnwidth]{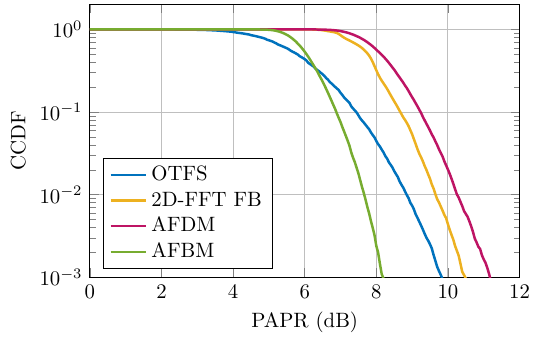}
    \vspace{-1.5em}
    \caption {\ac{CCDF} of the \ac{PAPR} for AFBM, AFDM, 2D-FFT FB, and OTFS.} 
    \label{fig:papr_ccdf}
\end{figure}

Fig.~\ref{fig:papr_ccdf} shows the \ac{CCDF} of the \ac{PAPR} for the considered waveforms. At a \ac{CCDF} level of $10^{-3}$, \ac{AFBM} attains a \ac{PAPR} of approximately 8 dB, which is about 1.7 dB, 2.3 dB, and 3 dB lower than that of \ac{OTFS}, 2D-FFT FB, and \ac{AFDM}, respectively, owing to its \ac{DAFT}-based
spreading.


\begin{figure}[htbp]
    \includegraphics[width=\columnwidth]{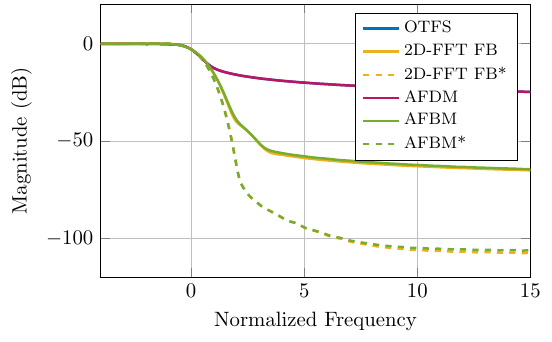}
    \vspace{-1.5em}
    \caption{\ac{PSD} versus normalized frequency for AFBM, AFDM, 2D-FFT FB, and OTFS. Asterisks indicate filter-bank waveforms using the PHYDYAS prototype filter, and unmarked waveforms use Hermite}
    \label{fig:psd}
\end{figure}

Next, Figure \ref{fig:psd} presents the \ac{PSD} of the considered waveforms. Filter bank-based modulations excel in spectral containment, showing an \ac{OOBE} gain of about 40 dB over OTFS and AFDM. Filters with even better \ac{PSD}
characteristics \cite{phydias} show an \ac{OOBE} gain of about 90 dB over OTFS and AFDM, they can be used both in \cite{AFBM} and \cite{2DFFT_FB} if some extra interference can be tolerated at the receiver, since exceeding the overlap factor threshold required for filter compensation introduces off-diagonal interference.

\begin{figure}[tb]
    \centering
    \includegraphics[width=\columnwidth]{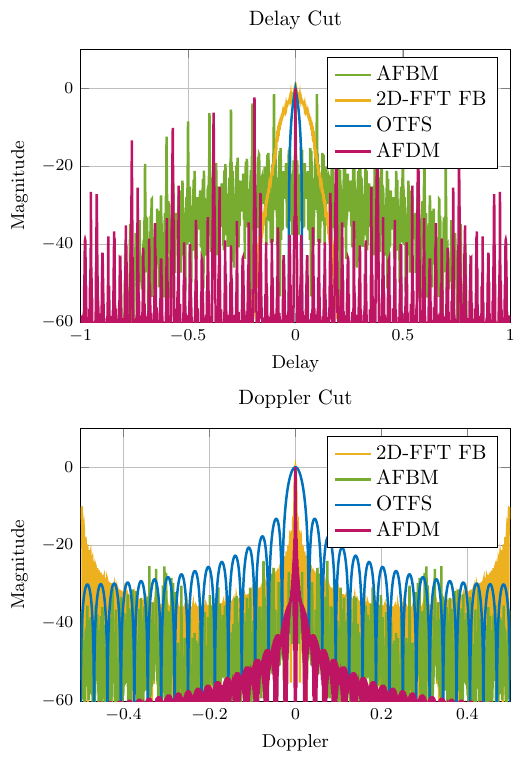}
    \vspace{-1.5em}
    \caption{Doppler and delay cuts of the \ac{AF} of each studied waveform, under a deterministic (all-ones) sensing symbol input.}
    \label{fig:af_cuts}
\end{figure}

\begin{figure}[tb]
    \centering
    \includegraphics[width=\columnwidth]{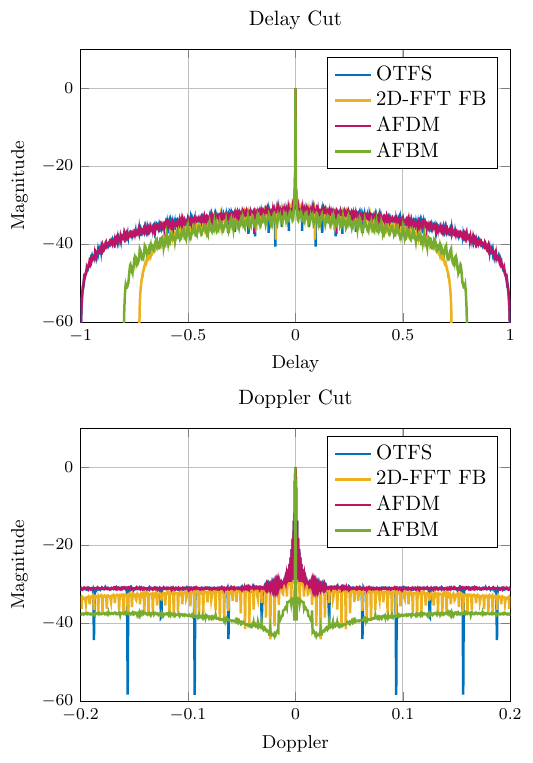}
    \vspace{-1.5em}
    \caption{Doppler and delay cuts of the \ac{AF} of each studied waveform, under a random \ac{QAM} sensing symbol input.}
    \label{fig:af_QAM_cuts}
\end{figure}

Fig.~\ref{fig:af_cuts} shows the AF cuts for a deterministic all-ones sensing symbol. In the delay cut, OTFS and 2D-FFT FB present a single unambiguous main lobe at the origin, with no sidelobes emerging above $-60$~dB. AFDM attains an equally sharp peak, but replicates it in a train of lobes reaching $\approx -25$~dB. AFBM shows the same peak superimposed on a dense correlation ridge that approaches $-5$~dB near the origin, the highest
delay ambiguity of the four. The ordering is reversed in the Doppler cut. AFDM
and AFBM concentrate their response in a one-bin main lobe, dropping immediately to a floor between $-20$ and $-40$~dB, while OTFS and 2D-FFT FB exhibit broader main lobes, with sidelobes decaying from $\approx -15$~dB. Under the random QAM sensing symbols of Fig.~\ref{fig:af_QAM_cuts}, the four
waveforms collapse onto a common response. A single-sample main lobe in
delay and a single-bin main lobe in Doppler, with AFBM reaching the lowest floor in the Doppler cut.
Thus, with data-carrying symbols the main lobes become practically indistinguishable and AFBM offers the best compromise.

The transmitter complexity of 2D-FFT FB is $K\big(\tfrac{L}{2} + L\log(L/2) + (L/2)\log(K) + N\log N + ON\big)$~\cite{2DFFT_FB}. For AFBM, it is $\mathcal{O}(ON + N\log N + N\log N + R\log R + 4L + 2L\log L + L)$~\cite{AFBM}, where the terms beyond the filter bank modulator account for the frequency-domain zero padding, the (I)DAFT operations and the compensation stage. Both schemes therefore share the filter bank cost $N\log N + ON$, given by an $N$-point IFFT and the prototype filter multiplications, and differ only in precoding, where the (I)DAFT operations of AFBM add complexity with respect to 2D-FFT FB. By contrast, OTFS's transmitter complexity is $\frac{MN}{2}\log_2(N)$ for a $M\times N$ lattice, and AFDM has a complexity of $\frac{E}{2}\log_2(E)+2E$, where the first term corresponds to the $E$-point IDFT and the second to the two diagonal chirp operations~\cite{Rou2024}.  Overall, the improved  characteristics provided by the filter-bank waveforms are obtained at the cost of up to three times higher complexity transmitter structure. Both 2D-FFT FB and AFBM require prototype-filter processing and multiple transform stages, whereas OTFS and AFDM rely primarily on FFT-based precoding.
\section{Conclusion}\label{sec:conclusion}

This paper compared 2D-FFT FB and AFBM, two ISAC filter bank waveforms differing only in precoding, using OTFS and AFDM as references. All four waveforms achieved BER within about 1 dB. The filter bank core yields about 40 dB lower OOBE than OTFS and AFDM, while the precoder drives the remaining differences. The DAFT spreading of AFBM gives the lowest PAPR, 1.7 to 3 dB below the other waveforms, and the sharpest Doppler cut of the AF, but higher delay sidelobes under deterministic sensing symbols, where 2D-FFT FB keeps a clean delay cut. With random QAM symbols these differences largely vanish and AFBM offers the best compromise, at up to three times the transmitter complexity of OTFS and AFDM. Future work will address hardware impairments in communications and sensing, together with channel estimation.

\balance

\ifCLASSOPTIONcaptionsoff
  \newpage
\fi

\bibliographystyle{IEEEtran}
\bibliography{bibtex/IEEEexample}

\end{document}

%% file: acro_list.tex
\newacronym{2DFFTFB}{2D-FFT FB}{two-dimensional FFT-precoded filter bank} 
\newacronym{RSU}{RSU}{roadside unit} 
\newacronym{CCDF}{CCDF}{Complementary cumulative distribution function} 
\newacronym{OFDM}{OFDM}{Orthogonal frequency division multiplexing} 
\newacronym{DS}{DS}{Doubly-selective} 
\newacronym{DFT-s-OFDM}{DFT-s-OFDM}{DFT-spread \ac{OFDM}} 
\newacronym{FBMC}{FBMC}{filter bank multi-carrier} 
\newacronym{GFDM}{GFDM}{generalized frequency division multiplexing} 
\newacronym{OCDM}{OCDM}{orthogonal chirp division multiplexing} 
\newacronym{AFBM}{AFBM}{affine filter bank modulation} 
\newacronym{AFDM}{AFDM}{affine frequency division multiplexing} 
\newacronym{OTFS}{OTFS}{orthogonal time frequency space} 
\newacronym{T-OTFS}{T-OTFS}{transcendentally-rotated OTFS} 
\newacronym{ODDM}{ODDM}{orthogonal delay-Doppler division multiplexing} 
\newacronym{SWHM}{SWHM}{sparse Walsh-Hadamard multiplexing} 
\newacronym{FMCW}{FMCW}{frequency modulated continuous wave} 
\newacronym{OTSM}{OTSM}{orthogonal time sequency modulation}

\newacronym{LCT}{LCT}{linear canonical transform}
\newacronym{ZT}{ZT}{Zak transform}
\newacronym{DZT}{DZT}{discrete Zak transform}
\newacronym{IDZT}{IDZT}{inverse discrete Zak transform}
\newacronym{FT}{FT}{Fourier transform}
\newacronym{IFT}{IFT}{inverse Fourier transform}
\newacronym{DFT}{DFT}{discrete Fourier transform}
\newacronym{IDFT}{IDFT}{inverse discrete Fourier transform}
\newacronym{AFT}{AFT}{affine Fourier transform}
\newacronym{DAFT}{DAFT}{discrete affine Fourier transform}
\newacronym{IDAFT}{IDAFT}{inverse discrete affine Fourier transform}
\newacronym{FFT}{FFT}{finite Fourier transform}
\newacronym{SFFT}{SFFT}{symplectic finite Fourier transform}
\newacronym{ISFFT}{ISFFT}{inverse symplectic finite Fourier transform}
\newacronym{HT}{HT}{Heisenberg transform}
\newacronym{WT}{WT}{Wigner transform}
\newacronym{frFT}{frFT}{fractional Fourier transform}
\newacronym{IfrFT}{IfrFT}{inverse fractional Fourier transform}
\newacronym{fnT}{fnT}{Fresnel transform}
\newacronym{IfnT}{IfnT}{inverse Fresnel transform}
\newacronym{LT}{LT}{Laplace transform}
\newacronym{ILT}{ILT}{inverse Laplace transform}
\newacronym{FB}{FB}{filter bank}
\newacronym{FP}{FP}{fractional programming}
\newacronym{LDT}{LDT}{Lagrangian dual transform}
\newacronym{QT}{QT}{quadratic transform}

\newacronym{A/D}{A/D}{analog-to-digital}
\newacronym{D/A}{D/A}{digital-to-analog}
\newacronym{TDL-A}{TDL-A}{tapped delay line model of type A}
\newacronym{ISAC}{ISAC}{integrated sensing and communications}
\newacronym{JCAS}{JCAS}{joint communications and sensing}
\newacronym{EM}{EM}{electromagnetic}
\newacronym{CP}{CP}{cyclic prefix}
\newacronym{B5G}{B5G}{beyond fifth generation}
\newacronym{3GPP}{3GPP}{$3^{\rm{rd}}$ generation partnership project} 
\newacronym{4G}{4G}{fourth generation}                                
\newacronym{5G}{5G}{fifth generation}                                 
\newacronym{6G}{6G}{sixth generation}
\newacronym{LTV}{LTV}{linear time-variant}
\newacronym{LTI}{LTI}{linear time-invariant}
\newacronym{LTVM}{LTVM}{linear time-variant multipath}
\newacronym{TV}{TV}{time-variant}
\newacronym{TI}{TI}{time-invariant}
\newacronym{1D}{1D}{one-dimensional}
\newacronym{2D}{2D}{two-dimensional}
\newacronym{3D}{3D}{three-dimensional}
\newacronym{NTN}{NTN}{non-terrestrial network}
\newacronym{LEO}{LEO}{low earth orbit}
\newacronym{IoT}{IoT}{internet-of-things}
\newacronym{mmWave}{mmWave}{millimeter-wave}
\newacronym{THz}{THz}{Terahertz}
\newacronym{V2X}{V2X}{vehicle-to-everything}
\newacronym{C-V2X}{C-V2X}{cellular-V2X}     
\newacronym{NR}{NR}{New Radio}              
\newacronym{ETSI}{ETSI}{European Telecommunications Standards Institute}  
\newacronym{EHF}{EHF}{extremely high-frequency}
\newacronym{BER}{BER}{bit-error-rate}
\newacronym{ICI}{ICI}{inter-carrier interference}
\newacronym{SotA}{SotA}{state-of-the-art}
\newacronym{DoF}{DoF}{degrees-of-freedom}
\newacronym{UAV}{UAV}{unmanned aerial vehicle}
\newacronym{CAV}{CAV}{connected autonomous vehicles}
\newacronym{MIMO}{MIMO}{multiple-input multiple-output}
\newacronym{SISO}{SISO}{single-input single-output}
\newacronym{UE}{UE}{user equipment}
\newacronym{AP}{AP}{access point}
\newacronym{CIR}{CIR}{channel impulse response}
\newacronym{CSI}{CSI}{channel state information}
\newacronym{PAR}{PAR}{perfect at receiver}
\newacronym{NAT}{NAT}{none at transmitter}
\newacronym{SNR}{SNR}{signal-to-noise ratio}
\newacronym{AWGN}{AWGN}{additive white Gaussian noise}
\newacronym{ML}{ML}{maximum likelihood}
\newacronym{OOB}{OOB}{out-of-band}
\newacronym{MX}{MX}{multiplexing}
\newacronym{DMX}{DMX}{demultiplexing}
\newacronym{AoA}{AoA}{angle-of-arrival}
\newacronym{AoD}{AoD}{angle-of-departure}
\newacronym{5GAA}{5GAA}{5G Automotive Association} 
\newacronym{6G-IA}{6G-IA}{6G Smart Networks and Services Industry Association}  
\newacronym{MUSIC}{MUSIC}{MUltiple SIgnal Classification}
\newacronym{ESPRIT}{ESPRIT}{estimation of signal parameters via rotational invariance techniques}
\newacronym{DP}{DP}{detection problem}
\newacronym{EP}{EP}{estimation problem}
\newacronym{KPI}{KPI}{key performance indicator}
\newacronym{ISI}{ISI}{inter-symbol interference}
\newacronym{TVIRF}{TVIRF}{time-variant impulse response function}
\newacronym{TVTF}{TVTF}{time-variant transfer function}
\newacronym{DVIRF}{DVIRF}{Doppler-variant impulse response function}
\newacronym{DVTF}{DVTF}{Doppler-variant transfer function}
\newacronym{DDSF}{DDSF}{delay-Doppler spread function}
\newacronym{OOBE}{OOBE}{out-of-band emission}
\newacronym{PAPR}{PAPR}{peak-to-average power ratio}
\newacronym{RCS}{RCS}{radar cross-section}
\newacronym{PIR}{PIR}{peak interference residual}
\newacronym{PSR}{PSR}{peak-to-sidelobe ratio}
\newacronym{PA}{PA}{power amplifier}
\newacronym{RF}{RF}{radio frequency}

\newacronym{CNIT}{CNIT}{Consorzio Nazionale Interuniversitario per le Telecomunicazioni}
\newacronym{PNRR}{PNRR}{Italian National Recovery and Resilience Plan}
\newacronym{DT}{DT}{discrete time}
\newacronym{CT}{CT}{continuous time}
\newacronym{UAVs}{UAVs}{unmanned aerial vehicles}
\newacronym{TF}{TF}{time-frequency}
\newacronym{DD}{DD}{delay-Doppler}
\newacronym{AFD}{AFD}{affine Fourier domain}
\newacronym{PHY}{PHY}{physical-layer}
\newacronym{FTN}{FTN}{faster-than-Nyquist}
\newacronym{SEFDM}{SEFDM}{spectrally efficient frequency division multiplexing}
\newacronym{HWIs}{HWIs}{hardware impairments}
\newacronym{PN}{PN}{phase noise}
\newacronym{CPE}{CPE}{common phase error}
\newacronym{MP}{MP}{message passing}
\newacronym{SIC}{SIC}{successive interference cancellation}
\newacronym{AF}{AF}{ambiguity function}
\newacronym{PMEPR}{PMEPR}{peak-to-mean envelope power ratio}
\newacronym{RRC}{RRC}{root-raised cosine}
\newacronym{DCP}{DCP}{double CP}
\newacronym{TBI}{TBI}{time-burst interference}
\newacronym{NBI}{NBI}{narrowband interference}
\newacronym{LO}{LO}{local oscillator}
\newacronym{CFO}{CFO}{carrier frequency offset}
\newacronym{AFS}{AFS}{affine Fourier series}
\newacronym{PSD}{PSD}{power spectral density}
\newacronym{SJ}{SJ}{sampling jitter}
\newacronym{CRBs}{CRBs}{Cram{\'e}r-Rao Bounds}
\newacronym{TX}{TX}{transmit}
\newacronym{RX}{RX}{receive}
\newacronym{TD}{TD}{time domain}
\newacronym{QAM}{QAM}{quadrature amplitude modulation}
\newacronym{RHS}{RHS}{right-hand side}
\newacronym{CFT}{CFT}{continuous-time Fourier transform}
\newacronym{SI}{SI}{self-interference}
\newacronym{IR}{IR}{impulse response}
\newacronym{LMMSE}{LMMSE}{linear minimum mean square error}
\newacronym{PAM}{PAM}{pulse amplitude modulation}
\newacronym{FD}{FD}{frequency domain}
\newacronym{SINR}{SINR}{signal-to-interference-plus-noise ratio}
\newacronym{SIR}{SIR}{signal-to-interference ratio}
\newacronym{PSK}{PSK}{phase-shift keying}
\newacronym{FIM}{FIM}{Fisher information matrix}
\newacronym{PDP}{PDP}{power delay profile}
\newacronym{FS}{FS}{Fourier series}
\newacronym{HWI}{HWI}{hardware impairment}
\newacronym{CRB}{CRB}{Cram\'{e}r-Rao Bound}
\newacronym{std}{std}{standard deviation}
\newacronym{CPP}{CPP}{chirp-periodic prefix}
\newacronym{SC}{SC}{Suppressed carrier}